\documentclass[%
 reprint,
 amsmath,amssymb,
 aps,
]{revtex4-2}

\usepackage{graphicx, float}
\usepackage[caption=false]{subfig} 
\usepackage{dcolumn}% Align table columns on decimal point
\usepackage{bm}
\usepackage{appendix}
\usepackage{afterpage}
\usepackage{natbib}
\usepackage{xcolor}
\usepackage{ulem}

\begin{document}

\title{Impact of particle production during inflation on the CMB detection}

\author{Xunliang Yang}
\affiliation{School of Fundamental Physics and Mathematical Sciences, Hangzhou Institute for Advanced Study, UCAS, Hangzhou 310024, China}
\affiliation{Institute of Theoretical Physics, Chinese Academy of Sciences,Beijing 100190,China}
\affiliation{University of Chinese Academy of Sciences, Beijing 100049, China}

\author{Zhe Yu}
\email{yuzhe@ihep.ac.cn}
\affiliation{Institute of Fundamental Physics and Quantum Technology, Department of Physics, School of Physical Science and Technology, Ningbo University, Ningbo, Zhejiang 315211, China}

\author{Zhoujian Cao}
\thanks {corresponding author: zjcao@amt.ac.cn}
\affiliation{School of Fundamental Physics and Mathematical Sciences, Hangzhou Institute for Advanced Study, UCAS, Hangzhou 310024, China}

\date{\today}% It is always \today, today,

\begin{abstract}
This work focuses on particle production described by a nonminimally coupled model during inflation. In this model, three parameters determine the characteristic frequency and strength of the induced gravitational waves (GWs). Considering the impact of particle production on inflation, we identify the parameter values that generate the strongest GWs without violating the slow-roll mechanism at the CMB scale. However, even with such extreme parameters, the power spectrum of induced GWs is only about $0.3\%$ of that of  vacuum GWs. This contribution remains insignificant when identifying the primary source of the detected CMB B-mode polarization. Furthermore, when our results are integrated with the constraints driven by P+ACT+LB+BK18, the contribution of induced GWs at CMB scales becomes negligible. In contrast, their impact on the scalar spectral index $n_s$ proves significant. For a range of parameter values, the Starobinsky inflation model yields predictions for $n_s$ that are consistent with the measurements obtained from P+ACT+LB+BK18.
\end{abstract}

\maketitle

\section{Introduction}
It has been a century since scientists first attempted to reconcile general relativity with quantum theory \cite{Carlip:2015asa,Kempf:2018gbt}. Nonetheless, in the absence of empirical data, a conclusive theoretical model has yet to emerge. The prevailing perspective holds that we cannot obtain such data from laboratory experiments because the test of quantum gravity (QG) needs a Milky Way-sized particle accelerator to touch the Planck length \cite{WOS:000404687900001}. However, during the inflationary epoch, the tensor perturbations (primordial gravitational waves) originate from quantum fluctuations of the gravitational field were amplified and resulted in a characteristic B-modes polarization signal in the Cosmic Microwave Background (CMB) \cite{Guzzetti:2016mkm}. This signal allows the direct detection of the inflationary epoch and potentially the quantum properties of gravity \cite{Guzzetti:2016mkm,WOS:001023808200001,Shiraishi:2016yun}. At present, a multitude of campaigns dedicated to the observation of the CMB B-mode polarization are actively underway (BICEP \cite{Moncelsi:2020ppj}, CLASS \cite{Dahal:2019xuf}, Simons Array \cite{POLARBEAR:2015ixw}, Simons Observatory \cite{SimonsObservatory:2018koc}, etc). In the near future, there will be more observational campaigns with higher precision trying to detect and measure the  B-mode polarization in the CMB (AliCPT \cite{WOS:000852408900003}, CMB-S4 \cite{CMB-S4:2020lpa}, etc).

After detecting CMB B-mode polarization, can we say that gravity possesses quantum properties? In fact, GWs can be sourced by any energetically-relevant components during inflation, and particles abundantly produced in the inflationary era would inevitably imprint signatures in primordial B-modes \cite{Shiraishi:2016yun}. Furthermore, various post-inflationary processes can generate gravitational waves. For example, a scale-invariant spectrum of gravitational radiation similar to that from inflation can arise from a continuous symmetry-breaking phase transition \cite{Krauss:1991qu}.  Gravitational waves can also be produced during the preheating or reheating phases due to explosive particle production, as well as from bubble collisions \cite{Caprini:2015zlo} or the loops arising throughout the cosmological evolution of the cosmic string network \cite{Wang:2022pav}, which produce a stochastic gravitational wave background. These gravitational waves can also affect the CMB B-mode polarization.

Accurately identifying the sources of CMB B-mode polarization is crucial for detecting quantum gravity and determining the energy scale of inflation. In order to distinguish the origins of CMB B-mode polarization, various methods have been proposed. For instance, the authors of \cite{Baumann:2009mq} defined causal $\tilde{B}$-modes and pointed out that $\tilde{B}$-mode correlations on angular scales $\theta>2^{\circ}$ are an unambiguous signature of inflationary tensor modes. The authors of \cite{Krauss:2010df} presented a new signature to discriminate between a spectrum of gravitational radiation generated by a self-ordering scalar field and that from inflation. One can also rule out many models through the constraints on scalar perturbations \cite{Guzzetti:2016mkm}.

As mentioned previously, scalar particle production during inflation can also result in GWs. Their contribution is typically insignificant compared to gravitational waves arising from quantum fluctuations of the gravitational field and has thus received limited attention. However, the authors of \cite{Yu:2023ity} introduced an extra scalar field $\chi$ that interacts with the inflaton $\phi$ via the nonminimally coupling and found it can generate GWs more efficiently. Therefore, it is essential to examine the impact of GWs in this scenario on the detection of the CMB B-mode polarization. Furthermore, the latest data release from the Atacama Cosmology Telescope (ACT) presents substantial changes in the constraints regarding the scalar spectral index $n_s$.
The results indicate that the constraints on $n_s$ driven by P+ACT+LB+BK18 joint analysis disfavors Starobinsky model within $ 1\sigma$\cite{ACT:2025fju,ACT:2025tim}. The presence of an additional scalar field introduces a correction to $n_s$, making it essential to evaluate its impact on $n_s$.

The structure of this paper is as follows. In Sec.~\ref{sec:II}, we provide a brief review of the model describing the nonminimal coupling between the inflaton $\phi$ and the scalar field $\chi$, and discuss the impacts of different parameters. In Sec.~\ref{sec:III}, we derive the appropriate parameters through numerical calculations and evaluate their impact on CMB detection. Finally, we devote Sec.~\ref{sec:IV} to the conclusion.

Throughout this paper we use units to make $c=\hbar=1$ and denote $M_{\mathrm{p}}\equiv1/\sqrt{8\pi G}$ as the Planck mass.
\section{Particle Production during Inflation}
\label{sec:II}
Our system is described by the following action \cite{Yu:2023ity}:
\begin{equation}
\begin{aligned}
    S &= \int \mathrm{d}^{4}x\sqrt{-g}\left[\frac{M_{{\mathrm{p}}}^{2}}{2}R - \frac{1}{2}\nabla^{\mu}\phi\nabla_{\mu}\phi - V(\phi) \right.\\
    &\left. - \frac{1}{2}\nabla^{\mu}\chi\nabla_{\mu}\chi - \frac{g^{2}}{2}(\phi-\phi_{0})^{2}\chi^{2} + \frac{1}{2}\xi R\chi^{2}\right].
\end{aligned}
\end{equation}
In this model, the scalar field $\chi$ interacts with the inflaton $\phi$ and Ricci scalar $R$ through the coupling term $\frac{g^2}{2} (\phi - \phi_0)^2 \chi^2$ and $\frac{1}{2}\xi R\chi^{2}$, where $g$ and $\xi$ are the coupling constants.\par
As $\phi$ gradually decreases and crosses $\phi_0$, the effective mass of the scalar field $\chi$ becomes zero.  At this point, the massless quantum field $\chi$ becomes unstable, and specific momentum modes of the field $\chi$ are excited. This result in the copious production of particles during inflation. The particles produced in this way act as a classical source for GWs.
The particle number of field $\chi$ produced during inflation depends on the choice of parameters $\phi_0,~g$ and $\xi$. Below, we briefly analyze the impact of these parameters on the induced GWs.\par
We start with the generation of particles. The equation of motion of mode functions $\chi_{\boldsymbol{k}}$ is given by:
\begin{equation}
    \ddot{\chi}_{\boldsymbol{k}}+3H\dot{\chi}_{\boldsymbol{k}}+\omega_{\boldsymbol{k}}^2\chi_{\boldsymbol{k}}=0,
\end{equation}
Taking the slow-roll approximation $\epsilon=-\dot{H}/H^2\ll 1 $, $\omega_{\boldsymbol{k}}^2$ is given by:
\begin{equation}
    \omega_{\boldsymbol{k}}^2\simeq\frac{k^2}{a^2}+g^2(\phi-\phi_0)^2-12\xi H^2.
\end{equation}

The condition for copious production of particles $\chi$ is $\omega_{\boldsymbol{k}}^2\leq0$. $\phi_0$ determines the time particle $\chi$ copiously produced and the characteristic frequency of the GWs, while $g$ and $\xi$ influence the amplitude and duration of particle production. Larger $g$ and smaller $\xi$ reduce the duration for which $\omega_{\boldsymbol{k}}^2 \leq 0$, thereby determining the strength of the GW spectrum. While this is a rough analysis, the actual impact of these parameters is more complex and will be discussed in the next section.
\begin{figure*}[t]
\centering
\begin{tabular}{c}
\includegraphics[width=0.8\textwidth]{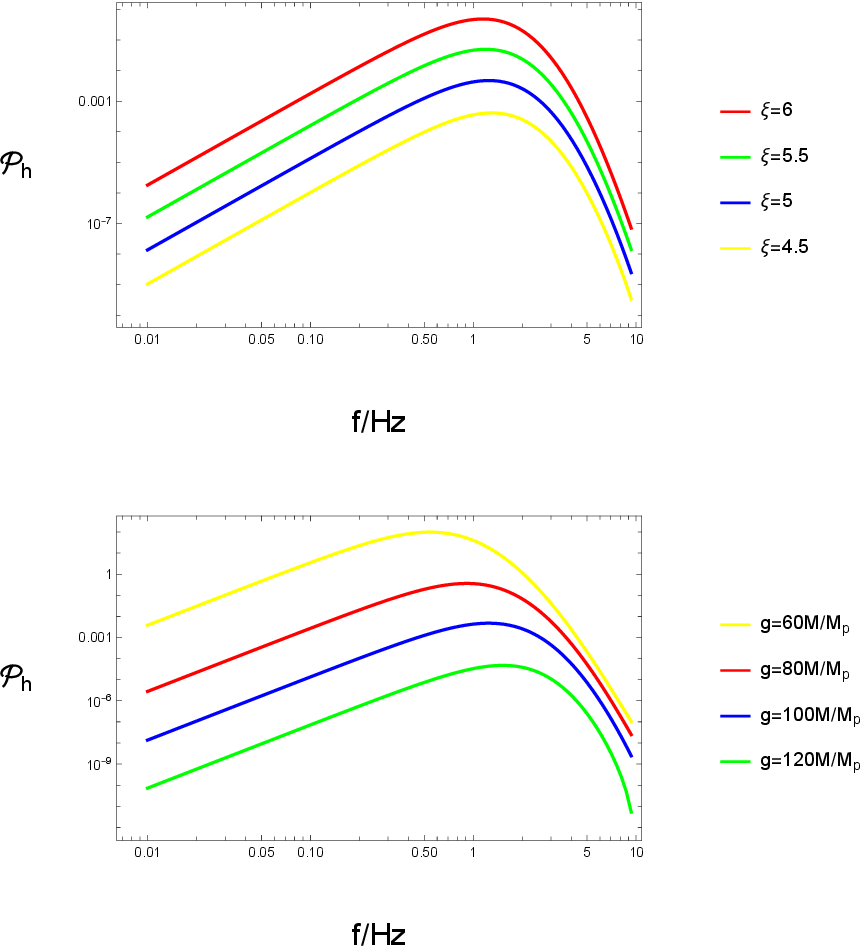}
\end{tabular}
    \caption{Power spectrum of the field $\chi$ for $\phi_0=4.57~\mathrm{M_p}$ when $\phi$ just cross $\phi_0$.
    \textbf{Top}:we fix $g=100~\mathrm{M/M_p}$ with different $\xi$. \textbf{Bottom}: we fix $\xi$=5 with different $g$. $p_0=a_0H_0$ denotes the scale exiting the horizon at the time when $\phi=\phi_0$. }
    \label{chi_ps1}
\end{figure*}

\begin{table*}[t]
\centering
\caption{The corresponding maximum value of $\xi$ for different g when $\phi_0=5.42~M_{\mathrm{p}}$.}
\label{g to xi}
\begin{tabular}{|c|c|c|c|c|c|c|c|c|c|c|}
\hline
$g~\mathrm{M/M_p}$  & 20 & 40 & 60 & 80 & 100 & 120 & 140  
& 160 &180 & 200 \\ \hline
$\xi$ & 0.57 & 0.94 & 1.27 & 1.58 & 1.86 & 2.14 & 2.4 
& 2.65 &2.89 & 3.13 \\ \hline
\end{tabular}
\end{table*}

\begin{figure*}[t]
\begin{tabular}{c}
\includegraphics[width=0.7\textwidth]{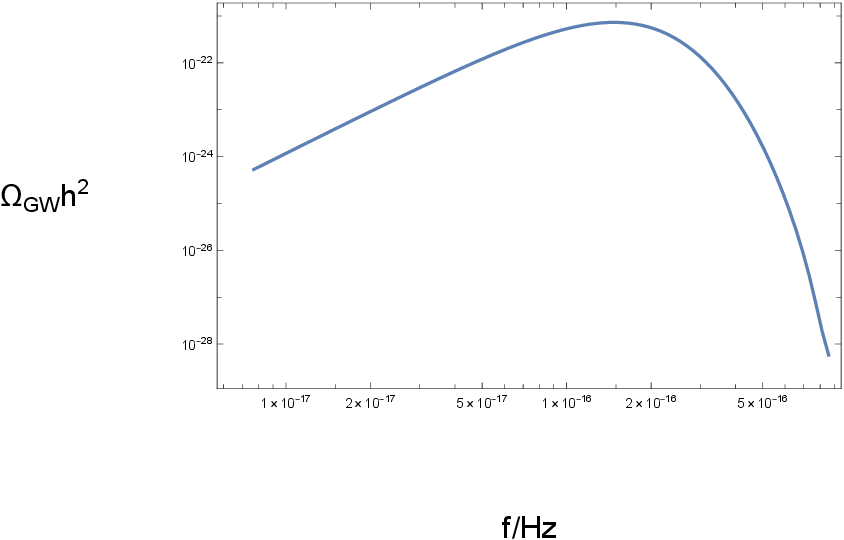}
\end{tabular}
    \caption{The energy spectrum of induced GWs with $\phi_0=5.42~\mathrm{M/M_p},g=160~\mathrm{M/M_p},\xi=2.65$.}
    \label{gwsps}
\end{figure*}

\begin{figure*}[t]
\begin{tabular}{c}
\includegraphics[width=0.7\textwidth]{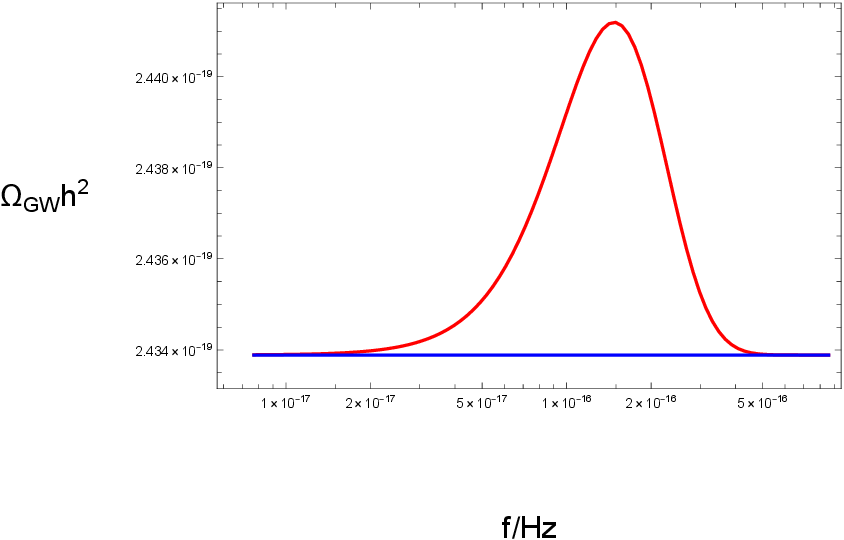}
\end{tabular}
    \caption{The total energy spectrum of GWs including the vacuum GWs and the particles induced one. The blue line represents the standard vacuum GWs predicted by the Starobinsky potential, and the red line is the total energy spectrum of GWs.}
    \label{totalps}
\end{figure*}
\section{Parameter Selection and Its Impact on detection of CMB}
\label{sec:III}

\subsection{Parameter Selection and Induced GWs}
In this section, we assess the impact of parameters through numerical calculations and preliminarily estimate the effect of particle production during inflation on the detection of CMB. The power spectrum of GWs is given by \cite{Yu:2023ity}
\begin{align}
    &\mathcal{P}_{h}(k)=\frac{2k^{3}}{\pi^{4}M_{\mathrm{p}}^{4}}\times\nonumber\\
    &\int_{0}^{\pi}\mathrm{d}\theta\sin^{5}\theta\int \mathrm{d}pp^{6}\left|\int^{\tau}\mathrm{d}\tau^{\prime}G_{k}(\tau,\tau^{\prime})\chi_{p}\chi_{{|\boldsymbol{k}-\boldsymbol{p}|}}\right|^{2}.
    \label{psgws}
\end{align}
The power spectrum of $\chi$ will feature a peak around $p_{\rm max}$ and the peak of the GW spectrum is near $p_{\rm max}$ \cite{Yu:2023ity}. Here, $a_0H_0$ represents the scale exiting the horizon when $\phi=\phi_0$.

For numerical calculations, we adopt the Starobinsky potential as a typical representative which can be expressed as \cite{Starobinsky:1980te,Yu:2023ity}:
\begin{equation}
    V(\phi)=M^2M_{\mathrm{p}}^2\left(1-\exp\left[-\sqrt{\frac{2}{3}}\frac{\phi}{M_{p}}\right]\right)^2,
\end{equation}
where $M=9.53 \times 10^{-6}~M_{\mathrm{p}}$. 
Among various inflationary models, scalar-field-based scenarios suffer from significant limitations, including the need for fine-tuning of initial conditions, sensitivity to the precise shape of the inflaton potential, and potential issues arising from quantum corrections. In contrast, f(R) gravity can naturally drive inflationary expansion in the early universe without exotic scalar potentials or
fine-tuned initial conditions \cite{Starobinsky:1980te,DeFelice:2010aj,Sotiriou:2008rp,Momeni:2025ylx}. The Starobinsky model is one of the earliest and most successful f(R) inflationary models, whose predictions are in excellent agreement with observational data. The Starobinsky model yields a scalar spectral index $n_s \approx 0.965$ and a tensor-to-scalar ratio $r \approx 0.004$ in perfect accord with the Planck 2018 and BICEP/Keck observations\cite{Planck:2015fie,BICEP:2021xfz}.In the next subsection, we will investigate the backreaction correction from the $\chi$ field to the potential, aiming to comply with the latest ACT observational constraints.

Based on numerical calculations , we plot the power spectrum of the $\chi$ field for $\phi_0 = 4.57~M_{\mathrm{p}}$ with varying values of $g$ and $\xi$ in Fig. \ref{chi_ps1}. From the results, it is evident that $p_{\rm max}$ is near $p_0 = a_0 H_0$, and specifically, larger $g$ and smaller $\xi$ yield higher $p_{\rm max}$. To match the CMB scale $k = 0.05~\mathrm{Mpc}^{-1}$, we choose $\phi_0 = 5.42~M_{\mathrm{p}}$ so that $p_0 = a_0 H_0 = 0.05~\mathrm{Mpc}^{-1}$.In \cite{Yu:2023ity}, the relation between the peak value of the power spectrum of the GWs induced by the produced particles and the power spectrum of the vacuum GWs is given by
\begin{equation}
P_h\simeq\frac{329.3\epsilon_0^{\frac{1}{4}}(\ln\gamma)^2}{g^{\frac{7}{2}}}\left(\frac{H}{M_{\mathrm{p}}}\right)^{\frac{3}
{2}}P_h^{(\mathrm{p})}.
\label{realtionph}
\end{equation}
$\mathcal{P}_h^{(p)}=(2H^2)/(\pi^2M_{\mathrm{p}}^2)$denotes the power spectrum of the vacuum GWs. This equation holds under the following conditions
\begin{equation}
\gamma \equiv \frac{H\Delta t}{2}=\frac{H^2\sqrt{2(6\xi+1)}}{-g\dot{\phi}_0}<\frac{1}{2}.
\end{equation}
Here, $\Delta t=t_e-t_i$ represents the duration of particle production, for $\phi_0 = 5.42~M_{\mathrm{p}}$, we get $\gamma\simeq 1.053$. That is to say Eq.~(\ref{realtionph}) does not hold under the conditions we are interested in. Therefore, detailed numerical calculations of the gravitational waves generated in this scenario are required.

In the following we investigate the parameters $\xi$ and $g$. First, we need to account for the backreaction of particle production on the evolution of the universe. As mentioned in Ref. \cite{Yu:2023ity}, excessive particle production during inflation can terminate inflation prematurely, failing to provide sufficient e-folding to resolve the flatness problem. In Appendix \ref{backreaction}, we present how the backreaction was considered and equations of perturbations. With backreaction, the effective potential is given by
\begin{align}
\label{Eff_Pot}
    V_{\rm eff} = V(\phi) + \frac{1}{2}g^2(\phi-\phi_0)^2\left\langle\chi^2\right\rangle.
\end{align}
The condition to prevent inflation from ending prematurely is:
\begin{align}
\label{veff>0}
 \frac{\mathrm{d}V_{\rm eff}}{\mathrm{d}\phi} =g^2\left(\phi-\phi_0\right)\left\langle\chi^2\right\rangle + \frac{\mathrm{d} V}{\mathrm{d} \phi} \geq 0,
\end{align}
From Eq. (\ref{veff>0}), we can derive an upper limit on particle produced during inflation.  In Table.~\ref{g to xi}, we present maximum values of $\xi$ for the given values of $g$. As expected, larger $g$ allows for larger $\xi$.

To estimate the effects on CMB B-Mode polarization, we need to identify the values of $g$ and $\xi$ which can induce the strongest GWs at the CMB scale. We calculate the energy spectrum of GWs at CMB scale $k = 0.05~\mathrm{Mpc}^{-1}$.  We find $g\approx160~M/M_{\mathrm{p}},\xi\approx2.65$ can induce the strongest GWs at CMB scale $k=0.05~\mathrm{Mpc}^{-1}$.

In Fig. \ref{gwsps} and Fig. \ref{totalps}, we present the resulting energy spectrum of the GWs signal and the total energy spectrum of GWs including the standard primordial one and the particles induced one. The spectrum features a peak around the CMB scale, consistent with our requirements. Through comparison with vacuum GWs, we reveal that even the peak value is only about $0.3\%$ of that of the vacuum GWs. Compared with the results in \cite{Yu:2023ity}, the gravitational waves obtained under both conditions exhibit similar single-peak waveforms, but their ratio to primordial gravitational waves differs by more than $10^4$.
The maximum peak value of the resulting GWs energy spectrum is on the order of $10^{-23}$, which is unobservable by any prospective GW experiment\cite{LISA:2017pwj,Ruan:2018tsw,TianQin:2015yph} or CMB B-Mode polarization experiment\cite{POLARBEAR:2015ixw,SimonsObservatory:2018koc,WOS:000852408900003,CMB-S4:2020lpa} and makes no detectable contribution to the effective number of neutrinos $\Delta N_{eff}$.\cite{Planck:2018vyg,Follin:2015hya,Hou:2011ec,Wallisch:2018rzj,Aich:2019obd}

\subsection{Impact on scalar spectral index}
In this section, we will study the effect of backreaction on the scalar spectral index and compare the results with observational data.

\begin{figure*}[t]
\begin{tabular}{c}
\includegraphics[width=0.67\textwidth]{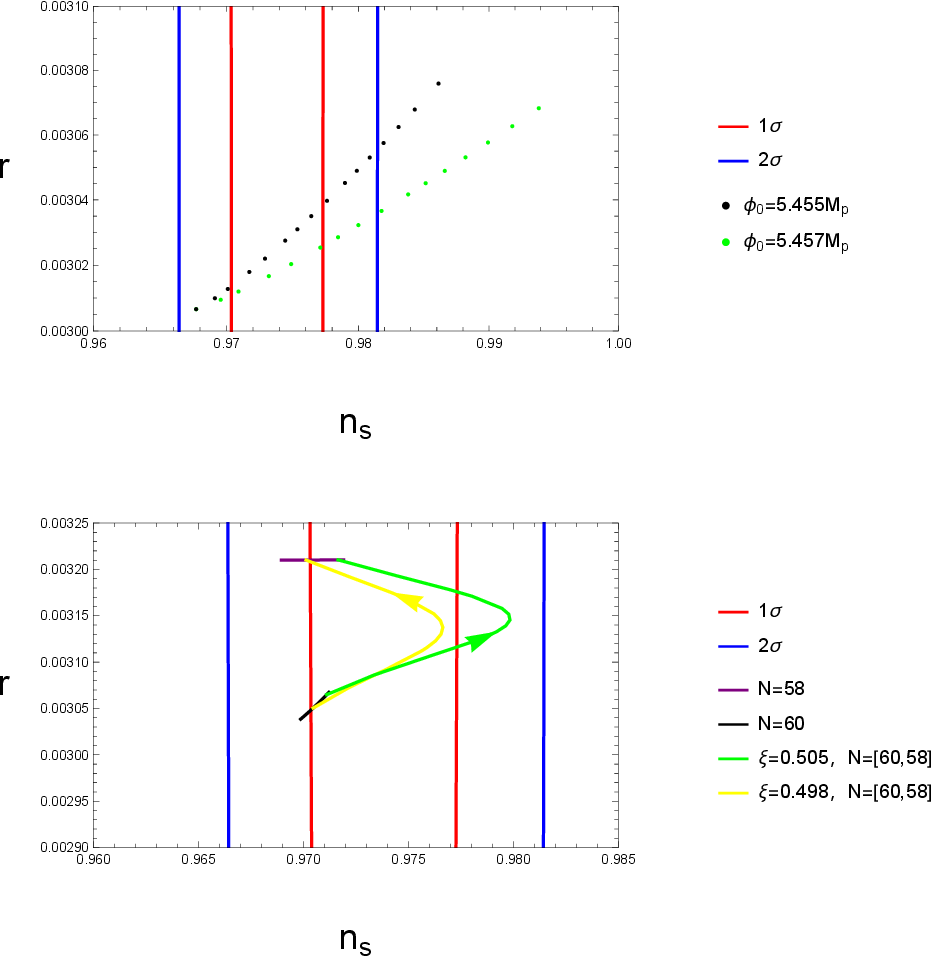}
\end{tabular}
    \caption{Prediction of our model with constraints on the scalar and tensor primordial power spectra according to ACT at $N=60$, shown in the r -$n_s$ parameter space. \textbf{Top}: We fix parameter $g=160~\mathrm{M/M_p}$ and change particle numbers through adjusting the value of $\xi$ from 1.4 to 2.65. \textbf{Bottom}:  We fix $\phi_0=5.5~\mathrm{M_p}, g=18~\mathrm{M/M_p}$ and change particle numbers through adjusting the value of $\xi$ from 0.49 to 0.505. The yellow and blue lines corresponds to the number of e-folding of inflation $N=[60,58]$ with $\xi=0.498,\xi=0.505$ respectively, where $N=[60,58]$ corresponds to the duration of correction of particle production.The arrow denotes the direction of evolution. }
    \label{nsn}
\end{figure*}

In Appendix \ref{backreaction}, we present the equations to calculate the first-order curvature perturbation $\mathcal{R}$. Utilizing these equations, we calculate the scalar spectral index $n_s$ and the tensor-to-scalar ratio $r$, finding $n_s=0.966541, r=0.00306727$ at $N=60$ and observe a significant increase in \( n_s \), which forms a bulge at $N < 60$ with the parameters we given earlier. The observed increase in the spectral index $n_s $ can be attributed to the production of particles. When $\phi$ crosses $\phi_0$, the effective potential becomes flat, resulting in an increase in $n_s$. To investigate this further, we select two representative values of $\phi_0$ and vary the particle numbers by adjusting $\xi$. In Fig. \ref{nsn}, we present the $n_s-r$ plane for various parameters along with the constraints on $n_s$ derived from the P+ACT+LB+BK18 results \cite{ACT:2025fju}. Notably, we find that certain parameter selections yield a scalar spectral index that falls within the $1~\sigma$ confidence region.

In the calculation for the top panel of Fig. \ref{nsn}, we found that the effects of $n_s$ still can only last for a short period. Therefore, we reduce the values of g to prolong the duration of particle production and modified \( \phi_0 \) to ensure that particle production occurs at the desired time. The results indicate that our model can significantly correct the scalar spectral index $n_s$ and the predictions of some values of parameters fall within the $1\sigma$ confidence region. Furthermore, after accounting for the constraints imposed by backreaction on the parameters, the influence of the produced particle can guarantee scalar spectral index $ n_s$ remains within the $1\sigma$ confidence region for approximately two e-folding numbers of inflation. The process of particle production is expected to persist for approximately one Hubble time, and the backreaction on the universe will sustain about 1-2 e-folds. Consequently, the spectral index $n_s$ is anticipated to increase by approximately 2-3 e-folds. However, in this scenario, the characteristic peak of the induced GWs shifts to larger scales, and the contribution of the produced particles to the total power spectrum of GWs at CMB scales becomes negligible. 

\begin{figure*}[t]
\begin{tabular}{c}
\includegraphics[width=0.67\textwidth]{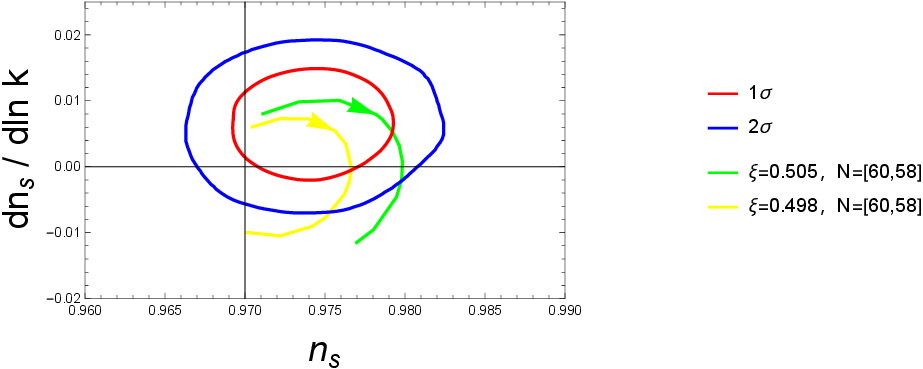}
\end{tabular}
    \caption{Prediction of the scalar spectral index $n_s$ and its running $\mathrm{d}n_s/\mathrm{d}\ln k$. The green and yellow lines correspond to the lines of identical colors in Fig. \ref{nsn}.}
    \label{dnsal}
\end{figure*}

In Fig. \ref{dnsal}, we show the prediction of the scalar spectral index $n_s$ and its running $\mathrm{d}n_s/\mathrm{d}\ln k$ with constraints on them derived from the P+ACT+LB result\cite{ACT:2025tim}. The particle number density rapidly decreases due to the expansion of the universe. This resulting in a sharp decline in $\mathrm{d}n_s/\mathrm{d}\ln k$. Therefore, at the end of particle production, the prediction of $\mathrm{d}n_s/\mathrm{d}\ln k$ lies outside the 2$\sigma$ confidence interval, which lasts about 1 e-fold.

The results indicate that, given the constraints on $n_s$ driven by P+ACT+LB+BK18, the particles generated through this process are unlikely to produce significant gravitational waves and will have a negligible correction on the tensor-to-scalar ratio $r$, on the order of $10^{-5}$. As such, it cannot be detected by any prospective CMB B-mode polarization experiment \cite{POLARBEAR:2015ixw,SimonsObservatory:2018koc,WOS:000852408900003,CMB-S4:2020lpa}. Therefore, we can ignore the effects of the produced scalar particles on CMB B-mode polarization.
The impact on $n_s$ is more worthy of attention. After accounting for the backreaction from the produced particles, $\phi$ will oscillate around $\phi_0$~\cite{Yu:2023ity} which could leave a noteworthy correction to the scalar spectral index \( n_s \). Since the order of the power spectrum remains unchanged, the tensor-to-scalar ratio \( r \) does not exhibit any observable variation. As shown in the bottom panel of Fig. \ref{nsn}, the predictions of the Starobinsky inflation model can return to within the $1\sigma$ confidence region given by ACT for certain parameter values. Obviously, such a phenomenon is not confined to this specific inflation model; thus, when comparing the predictions of various inflation models with observational data, it is essential to consider the production of particles during the inflationary phase.

\section{Conclusion and discussion}
\label{sec:IV}
In this paper, we examined the parameters governing particle production in the nonminimally coupled model and the GWs induced by the produced particles. The parameters $g$ and $\xi$, which serve as coupling constants in this model, determine the duration over which particles are copiously produced. The parameter $\phi_0$ sets the time at which this process begins and influences the characteristic frequency of the scalar field $\chi$. Then we verified these analyses by numerical calculations. Through the analysis of Eq.~(\ref{gwsps}), we infer that induced GWs will feature a peak near the characteristic frequency of $\chi$.
By selecting an appropriate value for $\phi_0$, we ensured that the induced GWs produced a peak close to the CMB scale, which allowed us to assess the impact of these particles on the detection of primordial GWs.

Taking into account the backreaction of particle production on inflation, we have identified specific values for $g$ and $\xi$ that generate the strongest GWs at the CMB scale without violating the slow-roll mechanism. However, even for extreme values of $g$ and $\xi$, the power spectrum of the induced GWs only reaches about $0.3\%$ of that of the vacuum GWs at their peak. It means the induced GWs are unobservable by any prospective GW experiment and CMB B-Mode polarization experiment.

When we take the constraints on $n_s$ driven by  P+ACT+LB+BK18 data set, we observed that the induced GWs become negligible. However, the impact of our model on $n_s$ is significant. It can correct the prediction of the Starobinsky inflation model back to fall within the $1\sigma$ confidence region driven by P+ACT+LB+BK18 result in some cases. This phenomenon can also be found in other inflation model, so it is significant to consider the impact on $n_s$ from particles produced during inflation when combining theoretical predictions with observational data.

\begin{acknowledgments}
This work was supported in part by the National Key Research and Development Program of China Grant No.~2021YFC2203001. Additionally, Zhe Yu is supported by the National Natural Science Foundation of China under Grant No.~12447176.
\end{acknowledgments}
\section*{Data Availability Statement}
Data sets generated during the current study are available from the corresponding author on reasonable request.
\appendix
\section{The basic equations with backreaction of perturbations}
In this section, we briefly review the basic equations with backreation of perturbations given by reference~\cite{Yu:2023ity} The spatially flat FRW metric in the conformal Newtonian gauge reads:
\label{backreaction}
\begin{align}
\label{Metric}
\mathrm{d} s^2=-\left(1+2\Psi\right)\mathrm{d}t^2+a^2\left[\left(1-2\Psi\right) \delta_{i j}+h_{i j}^{(1)}+\frac{h_{i j}}{2}\right]\mathrm{d} x^i \mathrm{d} x^j,
\end{align}
where the $\Psi$, $h_{i j}^{(1)}$ and $h_{ij}$ are the first-order scalar metric perturbation, the first-order tensor perturbation, and the second-order tensor perturbation, respectively.
The first-order curvature perturbation is given by
\begin{align}
    \mathcal{R}_{\boldsymbol{k}}=\Psi_{\boldsymbol{k}}+\frac{H\delta\phi_{\boldsymbol{k}}}{\dot\phi}.
\end{align}
Incorporating Hartree terms into the equation of motion, the background equations read:
\begin{widetext}
\begin{align} \label{Ein_Eq1_w_Br} 
3M_{\mathrm{p}}^2H^2 =& \frac{1}{1+\xi M_{\mathrm{p}}^{-2} \left\langle\chi^2\right\rangle}\left[\frac{1}{2}\dot{\phi}^2+\frac{1}{2}\left\langle\delta \dot{\phi}^2\right\rangle+\frac{1}{2}\left\langle\dot{\chi}^2\right\rangle+\frac{1}{2 a^2}\left\langle (\nabla\delta\phi)^2\right\rangle+\frac{1}{2 a^2}\left\langle(\nabla\chi)^2\right\rangle \right. \nonumber                               
\\
&\left. -6\xi H\left\langle \chi\dot\chi \right\rangle+ \frac{\xi}{a^2}\left\langle\nabla^2(\chi^2) \right\rangle+V(\phi)+\frac{1}{2}\frac{\partial^2 V }{\partial \phi^2}\left\langle\delta\phi^2 \right\rangle+\frac{1}{2}g^2\left(\phi-\phi_0\right)^2\left\langle\chi^2 \right\rangle \right],
\end{align}
\begin{align} \label{Ein_Eq2_w_Br}
 -M_{\mathrm{p}}^2\left(2\dot{H}+3H^2\right)= & \frac{1}{1+\xi M_{\mathrm{p}}^{-2} \left\langle\chi^2\right\rangle} \left[ \frac{1}{2} \dot{\phi}^2 +\frac{1}{2}\left\langle\delta \dot{\phi}^2\right\rangle +\left( \frac{1}{2} +2\xi\right)\left\langle\dot{\chi}^2\right\rangle - \frac{1}{6a^2}\left\langle(\nabla\delta\phi)^2\right\rangle  \right. \nonumber \\
& \left. - \frac{1}{6a^2} \left\langle(\nabla\chi)^2\right\rangle - \frac{2\xi}{3 a^2} \left\langle\nabla^2(\chi^2)\right\rangle + 4\xi H\left\langle \chi\dot\chi\right\rangle + 2\xi\left\langle \chi\ddot\chi\right\rangle - V(\phi) \right. \nonumber \\
&\left.- \frac{1}{2} \frac{\partial^2 V}{\partial \phi^2}\left\langle\delta \phi^2\right\rangle - \frac{1}{2} g^2\left(\phi-\phi_0\right)^2\left\langle\chi^2\right\rangle \right],
\end{align}
\begin{align} \label{KG_Eq_w_Br}
\ddot{\phi}+3 H \dot{\phi}+\frac{\partial V}{\partial \phi}+\frac{1}{2}\frac{\partial^3 V}{\partial \phi^3}\left\langle\delta \phi^2\right\rangle+g^2\left(\phi-\phi_0\right)\left\langle\chi^2\right\rangle=0, 
\end{align}
Then, the momentum-space linearized field perturbation equations are given by

 \begin{align}
\label{fo_pert_kG_phi}
\delta\ddot{\phi}_{\boldsymbol{k}}+3H \delta\dot{\phi}_{\boldsymbol{k}}+\Omega_{\boldsymbol{k}}^2 \delta\phi_{\boldsymbol{k}} =4\dot{\Psi}_{\boldsymbol{k}}\dot{\phi}-2\frac{\partial V}{\partial \phi}\Psi_{\boldsymbol{k}},
\end{align}
\begin{align}
\label{EoM_chi_k_w_Br}
\ddot{\chi}_{\boldsymbol{k}}+3H \dot{\chi}_{\boldsymbol{k}}+\omega_{\boldsymbol{k}}^2  \chi_{\boldsymbol{k}} =0,
\end{align}
with
\begin{align}
\Omega_{\boldsymbol{k}}^2=\frac{k^2}{a^2}+\frac{\partial^2 V}{\partial\phi^2}+g^2\left\langle\chi^2\right\rangle+\frac{1}{2}\frac{\partial^4 V}{\partial \phi^4}\left\langle\delta\phi^2\right\rangle,
\end{align}
\begin{align}\label{omega_k_chi}
\omega_{\boldsymbol{k}}^2=\frac{k^2}{a^2}+g^2\left(\phi-\phi_0\right)^2+g^2\left\langle\delta\phi^2\right\rangle-6\xi\left(\dot{H}^2+2H^2\right),
\end{align}
and the perturbed Einstein equations is
\begin{align}
\label{fo_pert_Ein_1}
3H\dot{\Psi}_{\boldsymbol{k}} + \left(  \frac{k^2}{a^2} + 3H^2 \right) \Psi_{\boldsymbol{k}} = - \frac{1}{2 M_\mathrm{p}^2} \left( \dot\phi\delta\phi_{\boldsymbol{k}} - \Psi_{\boldsymbol{k}}\dot\phi^2 + \frac{dV}{d\phi}\delta\phi_{\boldsymbol{k}} \right),
\end{align}
\begin{align}
\label{fo_pert_Ein_2}
\dot{\Psi}_{\boldsymbol{k}} + H \Psi_{\boldsymbol{k}} =\frac{\dot{\phi} }{2 M_\mathrm{p}^2} \delta \phi_{\boldsymbol{k}}.
\end{align}

\end{widetext}
Using  \eqref{fo_pert_Ein_1} and \eqref{fo_pert_Ein_2}, we can obtain $\Psi_{\boldsymbol{k}}$ as
\begin{align}
\label{psi_k}
\Psi_{\boldsymbol{k}}=\frac{\dot{\phi} \delta \dot{\phi}_{\boldsymbol{k}}+3H\dot{\phi} \delta \phi_{\boldsymbol{k}}+(\partial V/\partial \phi) \delta \phi_{\boldsymbol{k}}}{\dot{\phi}^2- 2 M_{\mathrm{p}}^2\left(k/a\right)^2},
\end{align}

The equation of motion for the first-order tensor perturbation $h_{i j}^{(1)}$ and second-order tensor perturbation $h_{ij}$ is given by
\begin{align}
h_{i j}^{(1)\prime\prime}(\tau,\boldsymbol{x})+2 \mathcal{H}h_{i j}^{(1)\prime}(\tau,\boldsymbol{x}) -\nabla^2 h_{i j}^{(1)}(\tau,\boldsymbol{x})=0,
\end{align}
\begin{align}
\label{EoM_IGW}
h_{i j}^{\prime\prime}(\tau,\boldsymbol{x})+2 \mathcal{H}h_{i j}^\prime(\tau,\boldsymbol{x}) -\nabla^2 h_{i j}(\tau,\boldsymbol{x})=\frac{4}{M_{\mathrm{p}}^2} \pi_{i j}^{TT}(\tau,\boldsymbol{x}),
\end{align}
Where a prime denotes the derivative with respect to the conformal time $\tau\equiv \int^t \mathrm{d}t/a$, and $\mathcal{H}\equiv a^{'}/a$ denotes the conformal Hubble parameter. The power spectrum, $\mathcal{P}_h = \sum_{\lambda=+,\times}k^3/(2\pi^2)|h_{\boldsymbol{k}}^\lambda|^2$, has the following form
\begin{align}
\label{PS_GW}
\mathcal{P}_h(k)=\frac{2 k^3}{\pi^4 M_{\mathrm{p}}^4} \int_0^{\pi} \mathrm{d} \theta \sin ^5 \theta \int \mathrm{d} p p^6\left|\int^{\tau} \mathrm{d} \tau^\prime G_{\boldsymbol{k}}\left(\tau, \tau^\prime\right)\chi_{\boldsymbol{p}} \chi_{|\boldsymbol{k}-\boldsymbol{p}|}\right|^2.
\end{align}
Where 

\begin{align}
h_{i j}= \sum\limits_{\lambda=+,\times} \int \frac{\mathrm{d} \boldsymbol{k}^3}{(2\pi)^{3/2}}\mathrm{e}^{\mathrm{i}\boldsymbol{k}\cdot\boldsymbol{x}}\mathrm{e}^\lambda_{i j}(\boldsymbol{k})h_{\boldsymbol{k}}^{\lambda}(\tau),
\end{align}
\begin{widetext}
\begin{align}
G_{\boldsymbol{k}}\left(\tau, \tau^\prime\right)= \frac{1}{k^3\tau^{\prime 2}}\left[\left(1+k^2 \tau \tau^\prime\right) \sin k(\tau-\tau^\prime)-k(\tau-\tau^\prime) \cos k(\tau-\tau^\prime)\right] \Theta(\tau-\tau^\prime).
\end{align}
\end{widetext}
\begin{align}
\pi_{i j}^{T T} 
&= \sum\limits_{\lambda=+,\times} \int \frac{\mathrm{d} \boldsymbol{k}^3 \mathrm{d} \boldsymbol{p}^3}{(2 \pi)^3} \mathrm{e}^{\mathrm{i} \boldsymbol{k} \cdot \boldsymbol{x}}\mathrm{e}^\lambda_{i j} e^{\lambda,l m} p_l p_m \chi_{ | \boldsymbol{k}-\boldsymbol{p} |} \chi_p.
\end{align}
\bibliography{main_}% Produces the bibliography via BibTeX.

\end{document}